\documentclass[aps,prl,twocolumn,superscriptaddress,amssymb,showpacs]{revtex4-1}
\usepackage{verbatim}
\usepackage{graphicx}
\usepackage{amsmath}
\graphicspath{{pict/}{}}

\newcommand\pictc[5]{\begin{figure}
                       \centerline{\vspace{0mm}\includegraphics[width=#1\columnwidth,height=0.7\textheight,keepaspectratio]{#3}}
                       \protect\caption{\protect\label{fig:#4} #5}\vspace{0mm}
                    \end{figure}            }
\newcommand\pict[4][1]{\pictc{#1}{!tb}{#2}{#3}{#4}}

\newcommand\rpict[1]{\ref{fig:#1}}

\newcommand\leqt[1]{\protect\label{eq:#1}}
\newcommand\reqtn[1]{\ref{eq:#1}}
\newcommand\reqt[1]{(\reqtn{#1})}

\begin{document}

\title{Parity-Time Anti-Symmetric Parametric Amplifier}

\author{Diana A. Antonosyan}
\author{Alexander S. Solntsev}
\author{Andrey A. Sukhorukov}
\email[]{Andrey.Sukhorukov@anu.edu.au}
\homepage[]{http://physics.anu.edu.au/nonlinear/ans}
\affiliation{Nonlinear Physics Centre, Research School of Physics and Engineering, The Australian National University, Canberra, ACT 2601, Australia}

\begin{abstract}
We describe the process of parametric amplification in a directional coupler of quadratically nonlinear and lossy waveguides, which belong to a class of optical systems with spatial parity-time (PT) symmetry in the linear regime. We identify a distinct spectral parity-time anti-symmetry associated with optical parametric interactions, and show that pump-controlled symmetry breaking can facilitate spectrally selective mode amplification in analogy with PT lasers. We also establish a connection between breaking of spectral and spatial mode symmetries, revealing the potential to implement unconventional regimes of spatial light switching through ultrafast control of PT breaking by pump pulses.

\end{abstract}


\maketitle

Light propagation in waveguiding structures with spatially distributed sections of loss and gain can be analogous to quantum wavepacket dynamics governed by a parity-time (PT) symmetric Hamiltonian~\cite{Guo:2009-93902:PRL}. Below a certain gain/loss level, such systems support PT-symmetric optical modes, which then exhibit the same average loss or gain~\cite{Ruschhaupt:2005-L171:JPA, El-Ganainy:2007-2632:OL}. However when gain or loss is increased, the PT-symmetry of modes breaks, and a mode with the strongest gain (or smallest loss) dominates, as demonstrated experimentally~\cite{Guo:2009-93902:PRL, Ruter:2010-192:NPHYS}. The phase transition associated with such PT-symmetry breaking opens new possibilities for light manipulation, such as PT-symmetric lasers~\cite{Feng:2014-972:SCI, Hodaei:2014-975:SCI}. Such lasers can achieve single-mode operation, where small difference in medium gain leads to a dramatic difference in mode amplification below and above the PT breaking threshold.

Parametric amplifiers are commonly used as an integral part of optical setups enabling flexible wavelength conversion and tunable signal gain, extending the range of lasers where gain media are limited to particular wavelengths. Wave amplification is efficiently realized in the regime of difference-frequency generation in media with quadratic optical nonlinearity~\cite{Boyd:2008:NonlinearOptics}. Importantly, the amplification rate is determined by the pump, enabling ultrafast all-optical tunability. In this work, we reveal the potential of PT-symmetric systems for optical parametric amplification, and identify a new regime of spectral PT anti-symmetry in such devices. Such devices can, on one hand, realize ultrafast spatial signal switching through pump-controlled breaking of PT symmetry, and on the other hand enable spectrally-selective mode amplification in analogy with PT lasers.

We consider a directional coupler composed of two waveguides in quadratically nonlinear medium, where modes exhibit different loss in each waveguide. It can be realized experimentally based on LiNbO$_3$ couplers where parametric gain was demonstrated~\cite{Schiek:2005-11109:APL}, as well as other platforms with $\chi^{(2)}$ nonlinearity.
The loss can be introduced, for example, by depositing a thin layer of metal on top of th waveguide~\cite{Guo:2009-93902:PRL}. An illustration of such structure with loss in one waveguide is presented in Fig.~\rpict{SchPTC}.
In the linear regime, at low light intensities, such coupler realizes PT-symmetric optical system~\cite{Guo:2009-93902:PRL}. However at higher intensities the effect of quadratic nonlinear interactions becomes important. It was predicted that in the regime of second-harmonic generation, when the signal and idler waves have identical spectra at half of the pump frequency, parametric interactions in PT couplers can support a rich variety of nonlinear modes~\cite{Li:2013-53820:PRA}. Furthermore, the formation of quadratic solitons in spatially extended PT-symmetric structures was analyzed in detail~\cite{Moreira:2012-53815:PRA, Moreira:2013-13832:PRA}. However, the fundamentally important regime of parametric amplification in PT systems remained unexplored.


We analyze the process of optical parametric amplification based on nonlinear mixing between strong pump, signal and idler waves, as illustrated in Fig.~\rpict{SchPTC}.
We model the wave propagation using coupled-mode equations~\cite{Boyd:2008:NonlinearOptics} in the regime of narrowband and undepleted pump,
\begin{equation} \leqt{CMEA}
 \begin{split}
    i \frac{\partial E_{s1}}{\partial z}
        &=-\beta_{s1} E_{s1}-i\gamma_{s1} E_{s1} - C_{s} E_{s2}+i \chi_1 E_{p1} E^{\ast}_{i1},\\
    i \frac{\partial E_{s2}}{\partial z}
        &=-\beta_{s2} E_{s2}-i\gamma_{s2} E_{s2} - C_{s} E_{s1}+i \chi_2 E_{p2} E^{\ast}_{i2},\\
    i \frac{\partial E_{i1}}{\partial z}
        &=-\beta_{i1} E_{i1}-i\gamma_{i1} E_{i1} - C_{i} E_{i2}+i \chi_1 E_{p1} E^{\ast}_{s1},\\
    i \frac{\partial     E_{i2}}{\partial z}
        &=-\beta_{i2} E_{i2}-i\gamma_{i2} E_{i2} - C_{i} E_{i1}+i \chi_2 E_{p2} E^{\ast}_{s2},\\
    i \frac{\partial E_{p1}}{\partial z}
        &=-\beta_{p1} E_{p1} - i\gamma_{p1} E_{p1} - C_{p} E_{p2},\\
    i \frac{\partial E_{p2}}{\partial z}
        &=-\beta_{p2} E_{p2} - i\gamma_{p2} E_{p2} - C_{p} E_{p1}.
 \end{split}
\end{equation}
Here the subscripts stand for signal (`s'), idler (`i'), and pump (`p') waves in two waveguides (`1' and `2'), $E$ are the mode amplitudes, $z$ is the propagation distance along the waveguides, $\beta$ are the propagation constants (for the pump mode, the propagation constant is adjusted to account for quasi-phase-matching through periodic poling of the ferroelectric domains~\cite{Boyd:2008:NonlinearOptics}),
$\gamma$ are the linear loss coefficients, $C$ are the mode coupling coefficients between the waveguides,
and $\chi$ are effective quadratic nonlinear coefficients.

\pict[0.6]{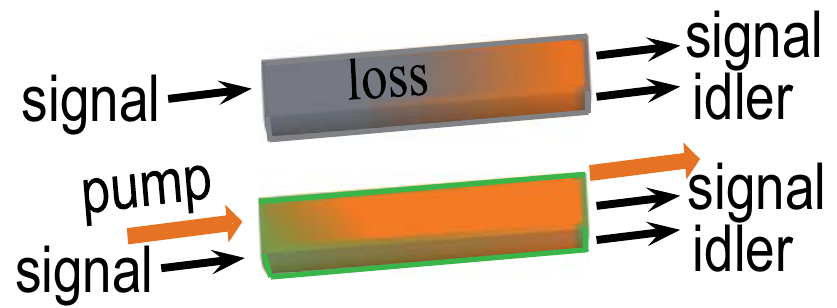}{SchPTC}{
Scheme of PT-symmetric nonlinear coupler with linear absorption in one waveguide.}

We find that PT-symmetry in the regime of parametric amplification can be achieved in the near-degenerate case when the waveguide parameters are practically the same at signal and idler frequencies, i.e. $\gamma_{i1}=\gamma_{s1} = \gamma_1$, $\gamma_{i2}=\gamma_{s2} = \gamma_2$, and $C_{s}=C_{i}=C$.
We assume that the waveguides are engineered such that $\beta_{s1} = \beta_{s2} = \beta_{s}$ and $\beta_{i1} = \beta_{i2} = \beta_{i}$, which is the regime required for linear PT-symmetry~\cite{Guo:2009-93902:PRL, Ruter:2010-192:NPHYS}.
Under usual experimental conditions the mode at higher pump frequency is much stronger localized compared to signal an idler~\cite{Solntsev:2014-31007:PRX}, leading to suppressed coupling between the waveguides and also very small sensitivity to metal deposited on top of waveguides, meaning that $C_p \equiv 0$ and $\gamma_{p1} = \gamma_{p2} = 0$.
We further consider a case of equal pump propagation constants in two waveguides, $\beta_{p1} = \beta_{p2} = \beta_p$.
The latter condition does not need to be satisfied if pump is coupled to one waveguide, which already enables full range of mode switching and amplification control as we demonstrate below.

To reveal the PT-symmetry properties of Eqs.~\reqt{CMEA}, we represent the equations for signal and idler waves in the Hamiltonian form,
\begin{equation} \leqt{HEM}
    i\frac{\partial\mathbf{a}}{\partial z}=\mathcal{{H}}\mathbf{a},
\end{equation}
where
\begin{equation}
\mathbf{a}(z)=
    \left( \begin{array}{cc}
        a_{s1}(z) \\
        a_{s2}(z) \\
        a^{\ast}_{i1}(z) \\
        a^{\ast}_{i2}(z)
    \end{array} \right) =
    \left( \begin{array}{cc}
        E_{s1}(z) e^{ -i (\beta + \beta_s) z }\\
        E_{s2}(z) e^{ -i (\beta + \beta_s) z } \\
        E^{\ast}_{i1}(z) e^{ i (\beta + \beta_i) z } \\
        E^{\ast}_{i2}(z) e^{ i (\beta + \beta_i) z }
    \end{array} \right) ,
\end{equation}
and
\begin{equation}\leqt{HF}
    \mathcal{H}=\left( \begin{array}{cccc}
    \beta-i\gamma_{1} & -C & iA_{1} & 0 \\
    -C & \beta-i\gamma_{2} & 0 & iA_{2} \\
    iA^{\ast}_{1} & 0 & -\beta-i\gamma_{1} & C \\
    0 & iA^{\ast}_{2} & C & -\beta-i\gamma_{2} \end{array} \right) ,
\end{equation}
where $\beta = (\beta_p - \beta_s - \beta_i) / 2$ defines the phase mismatch of parametric wave mixing, and $A_{1,2}$ are the normalized input pump amplitudes, $A_1 = \chi_1 E_{p1}(z=0)$ and $A_2 = \chi_2 E_{p2}(z=0)$.

A key result of our analysis is that the Hamiltonian possesses a {\em spectral anti-PT symmetry}, corresponding to a negative sign on the right-hand side of the following equality,
\begin{equation} \leqt{AntiSC}
    \mathcal{P}_{1,+}\mathcal{P}_{2,+}\mathcal{T}\mathcal{H}
        = -\mathcal{H}\mathcal{P}_{1,+}\mathcal{P}_{2,+}\mathcal{T}.
\end{equation}
Here $\mathcal{T}$ is a time-reversal operator which changes $z \rightarrow -z$ and performs a complex conjugation. The {\em parity operators operate in spectral domain}, interchanging the signal and idler waves,
\begin{equation}
 \mathcal{P}_{1,\pm}=\left\{ a_{s1}\leftrightarrow \pm a^{\ast}_{i1} \right\},
 \, \mathcal{P}_{2,\pm}=\left\{ a_{s2}\leftrightarrow \pm a^{\ast}_{i2} \right\}.
\end{equation}
We define the parity operators with both symmetric (`+') and antisymmetric (`-') transformations, since the latter will be useful in the following analysis. We note that such unusual symmetry is fundamentally different from the previously studied antisymmetric PT-metamaterials with modulated dielectric and magnetic properties~\cite{Ge:2013-53810:PRA}.

Since the Hamiltonian is linear in the undepleted pump regime, the dynamics of signal and idler waves is defined by the eigenmode solutions,
\begin{equation} \leqt{emodes}
  \mathbf{a}(z) = \widetilde{\mathbf{a}}(\sigma) \exp( i \sigma z  ),
\end{equation}
where $\sigma$ is an eigenvalue. The real part, ${\rm Re}\left(\sigma\right)$, defines the phase velocity, whereas the imaginary part determines the modal gain, $\Gamma = - {\rm Im}(\sigma)$.

\pict{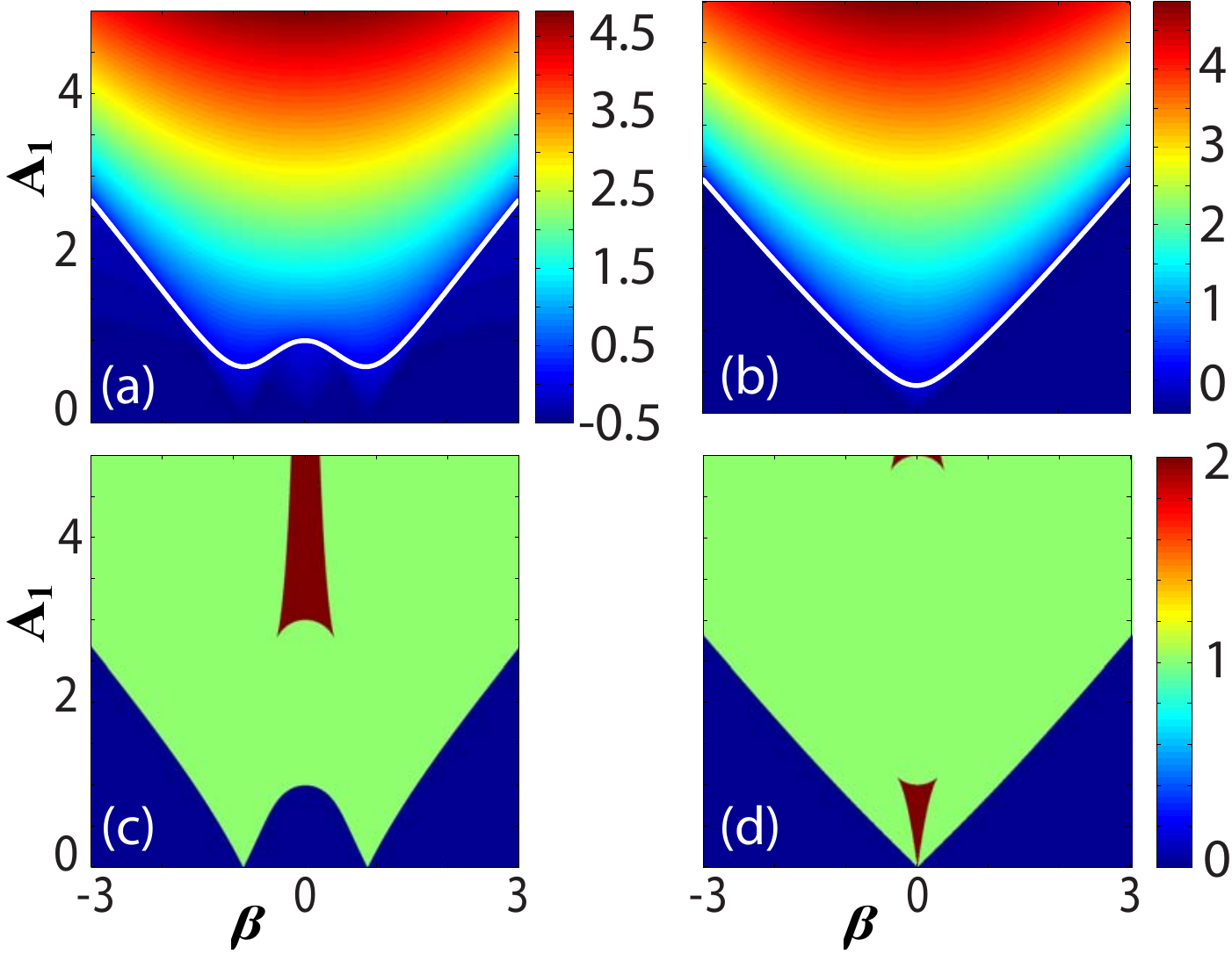}{SigmaABN}{
(a,b)~The largest mode gain (white line marks a zero gain level) and (c,d)~number of PT-symmetric mode pairs vs. the input pump in the first waveguide and the phase-mismatch.
The linear losses are (a,c)~$\gamma_{2s}=\gamma_{2i}=1$, (b,d)~$\gamma_{2s}=\gamma_{2i}=3$. For all the plots $\gamma_{1s}=\gamma_{1i}=0$, $C=1$ and $A_2 = 0$. }

\pict{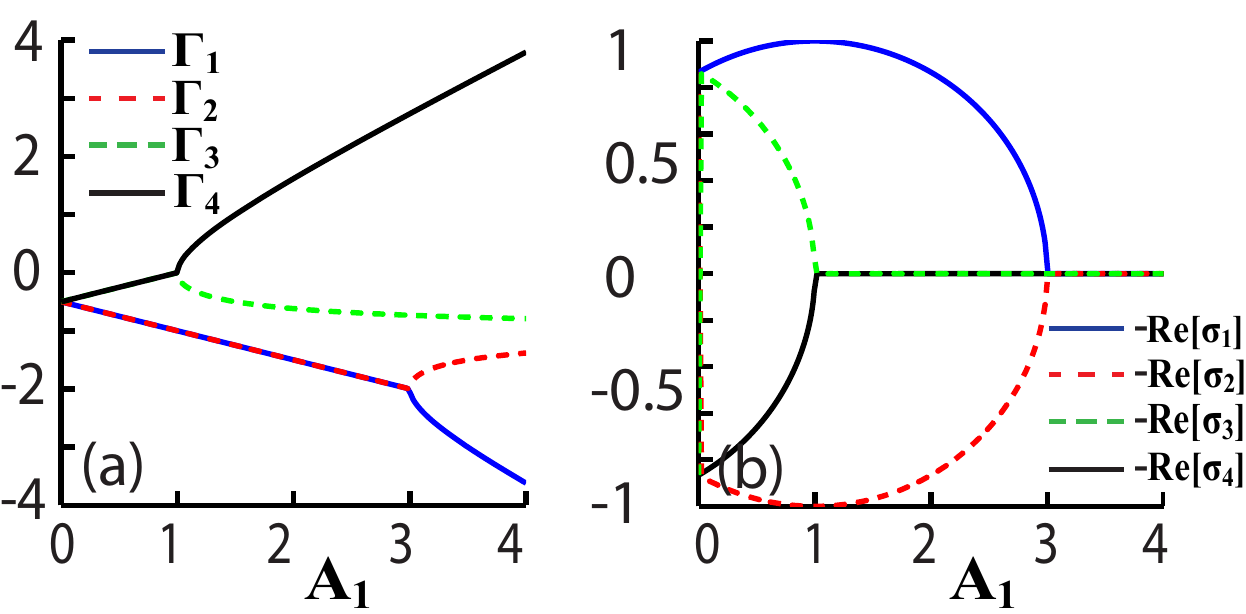}{RIS}{
Mode eigenvalues vs. the pump amplitude in the first waveguide ($A_{2}=0$): (a)~negative imaginary part defining mode amplification, $\Gamma = - {\rm Im}(\sigma)$, and (b)~real part defining propagation constant, ${\rm Re}(\sigma)$. Parameters are $\gamma_{1s}=\gamma_{1i}=0$, $\gamma_{2s}=\gamma_{2i}=1$, $C=1$, and $\beta=0$.
}

We determine the effect of spectral anti-PT symmetry on the eigenmode properties by substituting Eq.~\reqt{emodes} into Eq.~\reqt{HEM} and applying PT operator. We obtain that if $\widetilde{\mathbf{a}}(\sigma)$ is an eigenmode, then $\widetilde{\mathbf{a}}(-\sigma^\ast) = \mathcal{P}_{1,+}\mathcal{P}_{2,+}\mathcal{T} \widetilde{\mathbf{a}}(\sigma)$ is also an eigenmode. There are two possibilities how these relations can be satisfied. First, the mode can be PT-symmetric, when PT transformation maps the mode profile to itself (up to an overall phase), which happens if $-\sigma^\ast = \sigma$, and accordingly ${\rm Re}(\sigma) = 0$. Such {\em spectrally PT-symmetric modes would generally experience gain/loss different from other modes}, since there are no specific relations for the value of ${\rm Im}(\sigma)$.
Second, if the mode profile has broken PT-symmetry, the PT transformation relates two different modes with eigenvalues $\sigma_2 = - \sigma_1^\ast$. It follows that {\em a pair of spectrally PT-broken modes experience the same loss or gain}, ${\rm Im}(\sigma_1) = {\rm Im}(\sigma_2)$, but they have different phase velocities, ${\rm Re}(\sigma_1) = - {\rm Re}(\sigma_2)$. Remarkably, the established relations of mode symmetry and gain/loss
are reversed in comparison to previously studied spatial PT-symmetry in directional couplers~\cite{Guo:2009-93902:PRL, Ruter:2010-192:NPHYS}, due to the spectral {anti}-PT symmetry of parametric wave mixing.

We now demonstrate that the modal PT-breaking can be controlled by the pump beam. Due to the electronic nature of quadratic nonlinearity, such tuning can be ultrafast, directly following the pump profile in real time.
Overall, there are four eigenmodes of the Hamiltonian.
Accordingly, there can be three possible symmetry regimes: (i)~there are two mode pairs with broken PT-symmetry, (ii)~one pair of PT-broken modes and a pair of PT-symmetric modes, or (iii)~two pairs of PT-symmetric modes.
As an example, we present numerical analysis of the mode properties in a coupler with loss in the second waveguide, and pump coupled to the first waveguide.
We show the largest mode gain $\Gamma={\rm Max}[-{\rm Im}(\sigma)]$ in Figs.~\rpict{SigmaABN}(a,b) and the number of PT-symmetric mode pairs in Figs.~\rpict{SigmaABN}(c,d), depending on the pump amplitude $A_1$ and the phase-mismatch, for different values of linear loss.
We observe that in the regime when all modes have broken spectral PT symmetry [blue shaded regions in Figs.~\rpict{SigmaABN}(c,d)], the modes experience negative gain [c.f. Figs.~\rpict{SigmaABN}(a,b)]. This happens because pairs of eigenmodes exhibit the same amount of gain/loss, and effectively the amounts of gain and loss are averaged out between the eigenmodes. However upon transition to the region with spectrally PT-symmetric modes [green and red shaded regions in Figs.~\rpict{SigmaABN}(c,d)], there appears an unequal redistribution of gain and loss between the modes. One PT-symmetric eigenmode exhibits gain much larger then all other modes, while the latter experience stronger loss. Such sensitivity of amplification to PT-breaking threshold could be used to discriminate between multiple spectral modes, analogous to the concept of PT-lasers~\cite{Feng:2014-972:SCI, Hodaei:2014-975:SCI}.

Next, we establish a connection between the spectral PT-symmetry identified above, and the spatial wave dynamics due to waveguide coupling. We first consider a special case which can be treated analytically,
corresponding to perfect phase-matching, $\beta=0$, and additionally
${\rm Im}(A_{1}^\ast A_{2})=0$. The latter condition can be transformed to ${\rm Im}(A_{1}) = {\rm Im}(A_{2}) = 0$ under a substitution $\mathbf{a} \rightarrow \mathbf{a} \exp(i\varphi)$ with appropriately chosen constant phase $\varphi$.
Then, we find that
\begin{eqnarray}
 \mathcal{P}_{1,+}\mathcal{P}_{2,-} \mathcal{H} & = &
    \mathcal{H} \mathcal{P}_{1,+}\mathcal{P}_{2,-}, \leqt{Ppm} \\
 \mathcal{P}_{1,-}\mathcal{P}_{2,+} \mathcal{H} & = &
    \mathcal{H} \mathcal{P}_{1,-}\mathcal{P}_{2,+}.  \leqt{Pmp}
\end{eqnarray}
Accordingly, there appear two pairs of eigenmodes with $a^{\ast}_{i1}=\eta a_{s1}$ and $a^{\ast}_{i2}=-\eta a_{s2}$, where one pair with $\eta=+1$ has a profile symmetric with respect to $\mathcal{P}_{1,+}\mathcal{P}_{2,-}$ and second pair with $\eta=-1$ conforms to the symmetry $\mathcal{P}_{1,-}\mathcal{P}_{2,+}$. Remarkably, the signal dynamics of such modes is governed by equations resembling those for a linear PT-symmetric coupler~\cite{Guo:2009-93902:PRL, Ruter:2010-192:NPHYS},
\begin{equation} \leqt{CMsignal} 
    i \frac{\partial {\bf a}_{s}}{\partial z} = \mathcal{H}_r {\bf a}_{s},
  \, \mathcal{H}_{r}=\left( \begin{array}{cccc}
        i(\eta A_{1} -\gamma_{1})   & -C \\
       -C  & i(-\eta A_{2} -\gamma_{2}) \end{array} \right),
\end{equation}
where ${\bf a}_{s} = (a_{s1}; a_{s2})$.
We see that the linear loss is modified due to the parametric gain determined by the pump amplitudes. The coupler Eqs.~\reqt{CMsignal} are symmetric with respect to {\em spatial PT-symmetry}, up to a gauge transformation~\cite{Guo:2009-93902:PRL} expressed through the identity matrix ($\mathcal{I}$) in the following relation,
\begin{equation} \leqt{CouplerHPT}
    \mathcal{P}_{12}\mathcal{T} (\mathcal{H}_r - \bar{\rho}\; \mathcal{I}) = (\mathcal{H}_r - \bar{\rho}\; \mathcal{I}) \mathcal{P}_{12}\mathcal{T} ,
\end{equation}
where the {\em spatial parity operator} $\mathbf{P}_{12}$ swaps the mode amplitudes between the waveguides, $a_{s1} \leftrightarrow a_{s2}$ and $a_{i1} \leftrightarrow a_{i2}$. The coefficient $\bar{\rho} = (\eta A_{1} -\gamma_{1})/2 + (-\eta A_{2} -\gamma_{2})/2$ defines the average gain or loss between the two waveguides, which depends on the pump amplitudes and signal/idler mode symmetries corresponding to the different signs of `$\eta$'. The mode eigenvalues are
\begin{equation} \leqt{sigmaBeta0}
  \sigma = - i \bar{\rho} \pm
   (1/2) \sqrt{C^2 -[\gamma_1-\gamma_2 - \eta (A_1 + A_2)]^2} .
\end{equation}
We present characteristic dependencies of the eigenvalues on the pump amplitude in the first waveguide in Fig.~\rpict{RIS}.
We find that both {\em spatial and spectral PT-symmetry breaking occurs simultaneously} at the threshold
    $|\gamma_1-\gamma_2 - \eta (A_1 + A_2)| = C$.
However, the spatial and spectral symmetries are opposite: a mode pair is spatially PT-symmetric and has spectrally broken symmetry below threshold, whereas the situation is reversed above the threshold.
We show the evolution of the signal intensity along the waveguides is shown in Fig.~\rpict{AZ}.
At lower pump powers [Fig.~\rpict{AZ}(a)] signal periodically switched between the waveguides due to the beating between two modes exhibiting the same (negative) gain. However for stronger pump above the PT threshold [Figs.~\rpict{AZ}(b)], only one supermode with the strongest gain dominates and accordingly oscillations are suppressed.

\pict{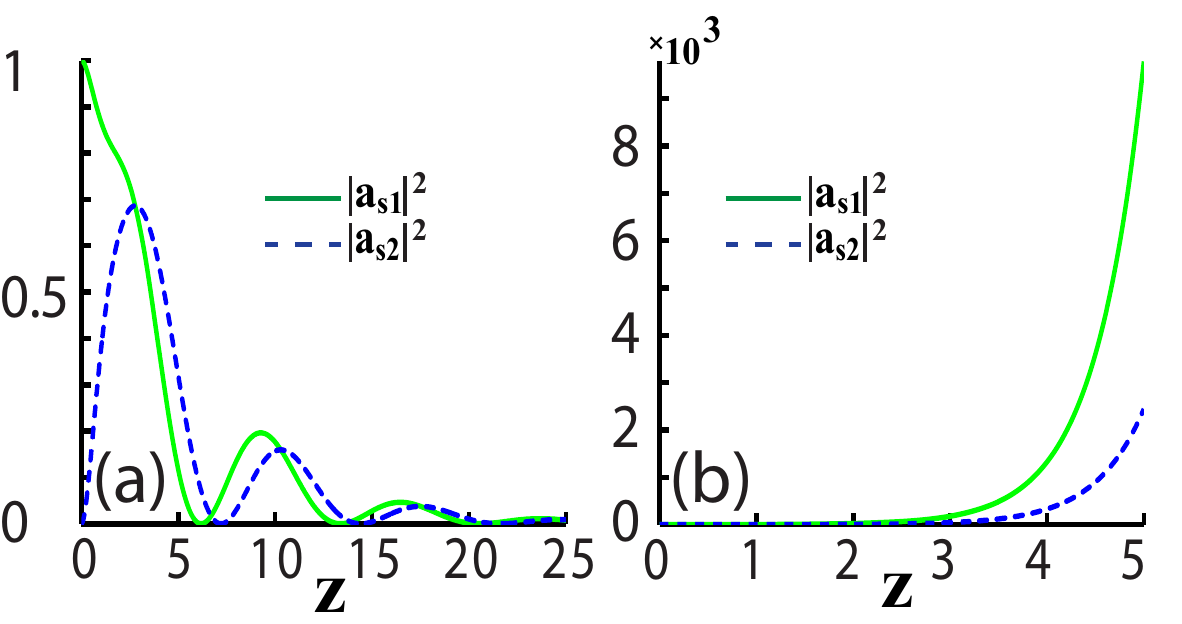}{AZ}{
Evolution of the signal mode intensity along the waveguides: solid (green) lines~--- the first waveguide without loss, dashed (blue) line~--- the second waveguide with loss. The pump amplitude is (a)~below PT-symmetry breaking threshold ($A_{1}=0.5$) and (b)~above threshold ($A_{1}=1.5$). Parameters are $\gamma_{1s}=\gamma_{1i}=0$, $\gamma_{2s}=\gamma_{2i}=1$, $C=1$, and $\beta=0$.
}

Finally, we investigate numerically a connection between spectral PT-breaking and spatial mode dynamics in a general case of non-zero phase mismatch. We present in Fig.~\rpict{A1B} the fraction of signal intensity in the first waveguide
depending on the phase-mismatch and the pump amplitude for different propagating distances.
We find that spatial dynamics strongly depends on the spatial PT-symmetry of modes in the linear regime (at low pump amplitudes $A_j \rightarrow 0$). If the linear modes are PT-symmetric, $|\gamma_1-\gamma_2| < C$, then increase of pump amplitude can control the period of mode coupling between the waveguides, while the oscillations get suppressed close to the spectral PT threshold, see Figs.~\rpict{A1B}(a,c) and Fig.~\rpict{SigmaABN}(c).
However if the linear modes are PT-broken, $|\gamma_1-\gamma_2| > C$, then mode beating between the waveguides still occurs below the spectral PT threshold, but with very small modulation amplitude, see Figs.~\rpict{A1B}(b,d) and Fig.~\rpict{SigmaABN}(d). In all cases, the mode beating changes faster with pump amplitude at longer distances.

\pict{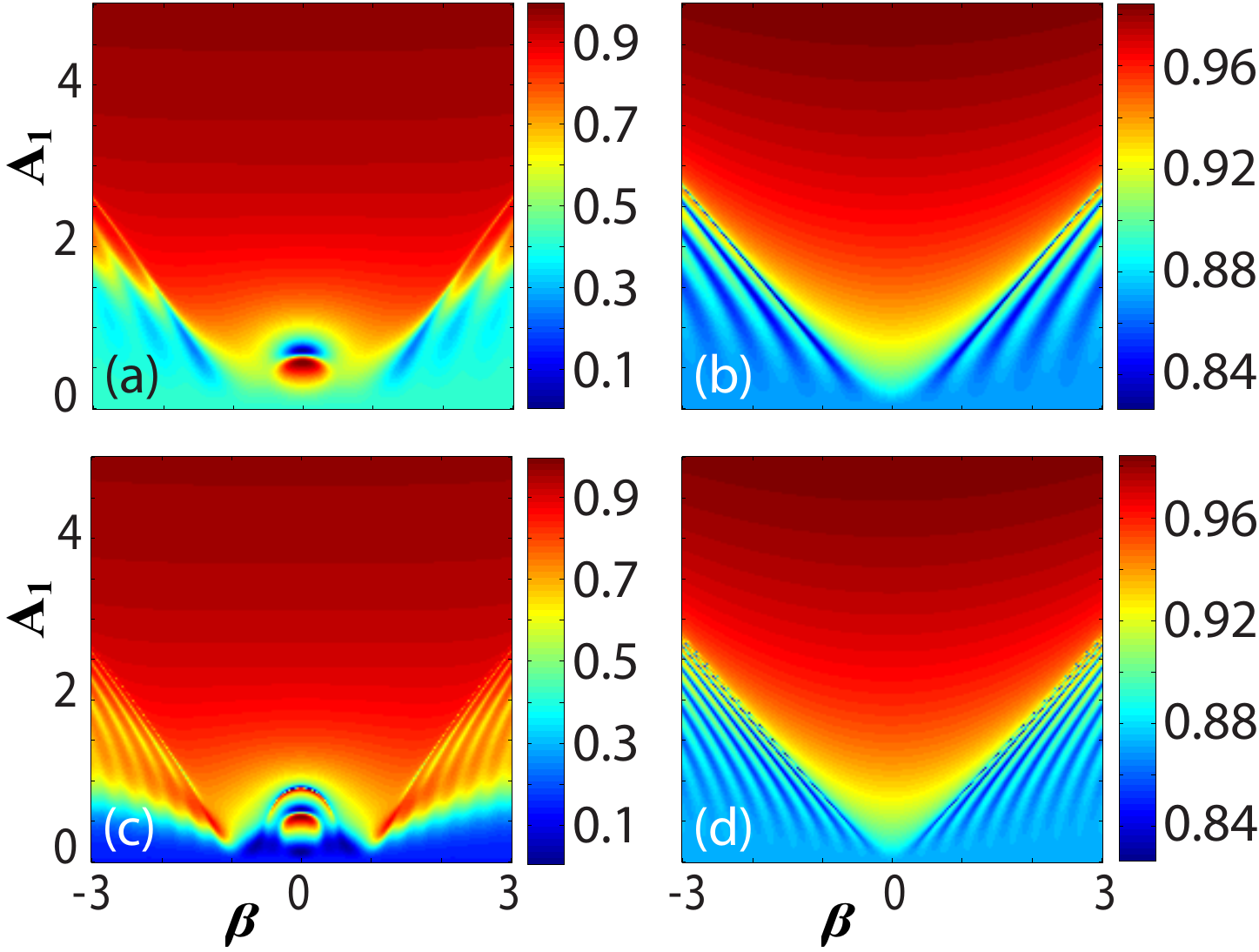}{A1B}{
Fraction of signal intensity in the first waveguide, $|a_{s1}|^{2}/(|a_{s1}|^{2}+|a_{s2}|^{2})$,
vs. the input pump in the first waveguide and the phase-mismatch. Propagation distances are (a,b)~$L=5C$ and (c,d)~$L=10C$.
Parameters correspond to (a,c)~Figs.~\rpict{SigmaABN}(a,c) and (b,d)~Figs.~\rpict{SigmaABN}(b,d).
}

In conclusion, we have identified an anti-PT spectral symmetry of a parametric amplifier based on a quadratically nonlinear coupler with different losses in two waveguides. For pump powers below the threshold, the modes form pairs with broken PT symmetry and same gain/loss, whereas above the threshold one PT-symmetric mode experiences the largest gain. This can facilitate spectrally-selective mode amplification, and we expect that this effect will become even more pronounced in resonator structures in analogy with single-mode operation of PT microring lasers.
We have further established an underlying connection between spectral anti-PT and conventional spatial PT mode symmetries, which reveals the possibility to control spatial light switching and amplification through parametric gain. Accordingly, the suggested platform can implement various unconventional regimes of light control previously suggested for PT-symmetric structures, such as unidirectional~\cite{Bender:2013-234101:PRL} and nonreciprocal~\cite{Peng:2014-394:NPHYS} operation, but with the advantage of ultrafast all-optical control of PT-symmetry regimes by pump pulses.
We anticipate that, due to the universality of parametric amplification processes,
these concepts will be extended to different physical mechanisms including Kerr-type optical nonlinearity, as well as a broad range of other systems including
cold atomic Bose-Einstien condensates~\cite{Vogels:2002-20401:PRL} in engineered potentials~\cite{Graefe:2012-444015:JPA, Cartarius:2012-13612:PRA} and exciton-polariton condensates~\cite{Lien:2015-24511:PRB}.

This work was supported by the Australian Research Council, including Discovery Project DP130100135 and Future Fellowship FT100100160.

\end{document}